\documentclass{article}
\PassOptionsToPackage{varqu}{inconsolata}
\usepackage{colm2024_conference}

\usepackage[utf8]{inputenc}
\usepackage[T1]{fontenc}

\usepackage{amsmath,amsfonts,bm}

\def\eqref#1{equation~\ref{#1}}

\def\1{\bm{1}}

\DeclareMathAlphabet{\mathsfit}{\encodingdefault}{\sfdefault}{m}{sl}
\SetMathAlphabet{\mathsfit}{bold}{\encodingdefault}{\sfdefault}{bx}{n}

\usepackage{mathtools}                 

\usepackage{amsmath}
\usepackage{amssymb}
\usepackage{booktabs}
\usepackage{graphicx}
\graphicspath{{figures/}}
\usepackage{microtype}
\usepackage{multirow}
\usepackage{makecell}
\usepackage{xcolor}
\usepackage{xspace}
\usepackage{hyperref}
\usepackage{url}
\usepackage{tcolorbox}
\usepackage{colortbl}
\tcbuselibrary{skins}

\newcommand{\myparagraph}[1]{\noindent\textbf{#1}}

\newcommand{\method}{\textsc{TCR}\xspace}

\newcommand{\website}{}
\newcommand{\projectpage}{https://github.com/DAGroup-PKU/Temporal-Context-Routing}

\definecolor{absbg}{RGB}{247,241,241}
\definecolor{absframe}{RGB}{140,21,21}
\definecolor{abstitle}{RGB}{140,21,21}
\definecolor{sectioncolor}{RGB}{153,46,46}
\definecolor{qwenpurple}{RGB}{92,59,218}

\hypersetup{colorlinks=true, citecolor=sectioncolor, linkcolor=sectioncolor, urlcolor=sectioncolor}

\newtcolorbox{thesisbox}[1]{
  enhanced,
  boxrule=0pt,
  frame hidden,
  borderline west={3pt}{0pt}{violet!65!black},
  colback=violet!4,
  sharp corners,
  left=9pt,
  right=9pt,
  top=7pt,
  bottom=7pt,
  before skip=7pt,
  after skip=7pt,
  before upper={
    \textbf{\color{violet!65!black}#1}\par\smallskip
  },
}

\makeatletter
\def\section{\@startsection {section}{1}{\z@}{-2.0ex plus
    -0.5ex minus -.2ex}{1.5ex plus 0.3ex
minus0.2ex}{\large\bf\color{sectioncolor}\raggedright}}
\def\subsection{\@startsection{subsection}{2}{\z@}{-1.8ex plus
-0.5ex minus -.2ex}{0.8ex plus .2ex}{\normalsize\bf\color{sectioncolor}\raggedright}}
\makeatother

\fancypagestyle{firstpage}{
  \fancyhead{}
  \lhead{%
    \raisebox{-9pt}{\includegraphics[height=38pt]{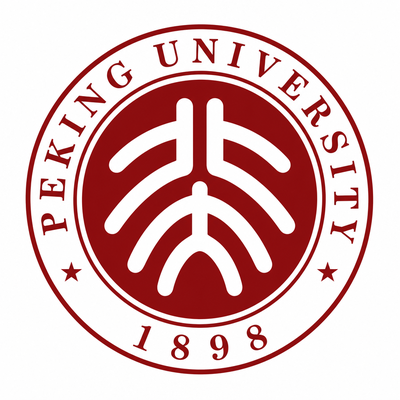}}%
    \hspace{10pt}%
    \raisebox{-5pt}{\includegraphics[height=32pt]{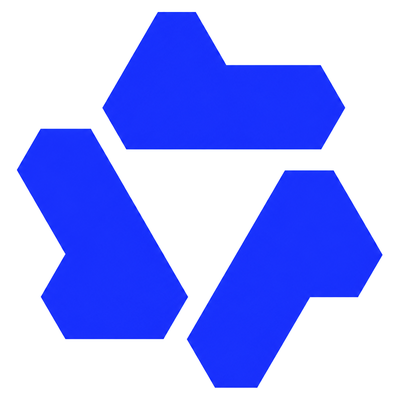}}%
  }
  \rhead{}
  \renewcommand{\headrulewidth}{2.2pt}
  \renewcommand{\headrule}{%
    {\color{abstitle}\hrule height \headrulewidth width \headwidth}%
    \vskip-\headrulewidth
  }
}

\makeatletter
\def\@maketitle{\vbox{\hsize\textwidth
 \kern-0.5cm
 {\centering {\Large\bf \@title\par}}
 \vspace{0.41cm}
 {\color{abstitle}\hrule height 0.7pt width \textwidth}
 \vspace{0.26cm}
 \begin{quote}\rule{\z@}{6pt} {\centering\par{\@author}\par}\end{quote}%
 \vskip 0.25in minus 0.1in
}%
\begin{center}
\vspace{-1cm}
\ifdefempty{\website}{}{\homepage~~\href{\website}{\textbf{Project Page: }\texttt{\website}}}
\end{center}
\thispagestyle{firstpage}}
\makeatother

\newtcolorbox{coverabstract}{
  enhanced,
  colback=absbg,
  colframe=absframe,
  boxrule=0.75pt,
  arc=6pt,
  left=14pt,
  right=14pt,
  top=10pt,
  bottom=10pt,
  before skip=6pt,
  after skip=10pt,
}

\renewenvironment{abstract}{%
  \begin{coverabstract}
  \centerline{\color{abstitle}\large\bfseries Abstract}%
  \vspace{0.7ex}%
  \noindent\ignorespaces
}{%
  \par\vspace{0.85em}
  \noindent \textbf{Project page:} \url{\projectpage}%
  \end{coverabstract}
}

\title{The Missing Temporal Link: Temporal Context Routing \\for Script-Driven Audio-Video Generation}

\author{
\makebox[\linewidth][c]{%
\begin{minipage}{\textwidth}
\centering
{\large\bfseries
Yichen Liu\textsuperscript{1}\quad
Quanwei Zhang\textsuperscript{2}\quad
Haozhe Wang\textsuperscript{3}\quad
Donghao Zhou\textsuperscript{4}}\\[3pt]
{\large\bfseries
Jiankun Zhang\textsuperscript{5}\quad
Xiaojie Li\quad
Yang Shi\textsuperscript{2}\quad
Jiaming Liu\textsuperscript{2,*}\quad
Ruihua Huang\textsuperscript{2}}\\[3pt]
{\large\bfseries
Yingtian Zou\textsuperscript{6}\quad
Daquan Zhou\textsuperscript{1,*}}\\[7pt]
{\small
\textsuperscript{1}PKU\quad
\textsuperscript{2}Qwen Applications\quad
\textsuperscript{3}HKUST\quad
\textsuperscript{4}CUHK\quad
\textsuperscript{5}UChicago\quad
\textsuperscript{6}SJTU}\\[3pt]
{\small\textsuperscript{*}Corresponding authors}
\end{minipage}}
}

\begin{document}

\maketitle

\begin{abstract}
Joint audio-video generation models have made substantial progress in visual quality and audio-visual synchronization. However, they still provide limited control over when shot transitions occur and dialogue is spoken. This limitation constrains their application in script-driven content creation, where timing errors can undermine narrative coherence and the viewing experience. Current joint generators align video and audio representations on a shared temporal axis, yet the precise timing of shots and dialogue specified in a structured prompt is encoded only in the prompt’s text representation and remains unaligned with the temporal coordinates of either modality. Consequently, video and audio may remain synchronized with each other while both fail to follow the script timeline. This mismatch motivates us to extend temporal alignment beyond video and audio to include the structured script. We therefore introduce \textbf{Temporal Context Routing (TCR)}, which maps the script timing onto the shared temporal axis of video and audio generation and routes each prompt’s guidance to the corresponding positions in both modalities. Compared with the baseline on 200 test scripts, TCR \textbf{reduces Shot Boundary MAE by 96\%, from 1.11 s to 0.042 s}, and \textbf{raises Dialogue Acc@0.5 s from 28.3\% to 84.1\%}. TCR achieves these improvements while maintaining visual quality and audio-visual synchronization comparable to those of the baselines. A user study further shows that participants prefer TCR on all five evaluated dimensions.

\end{abstract}

\section{Introduction}
\label{sec:introduction}

\begin{figure*}[t]
  \centering
  \includegraphics[width=\textwidth]{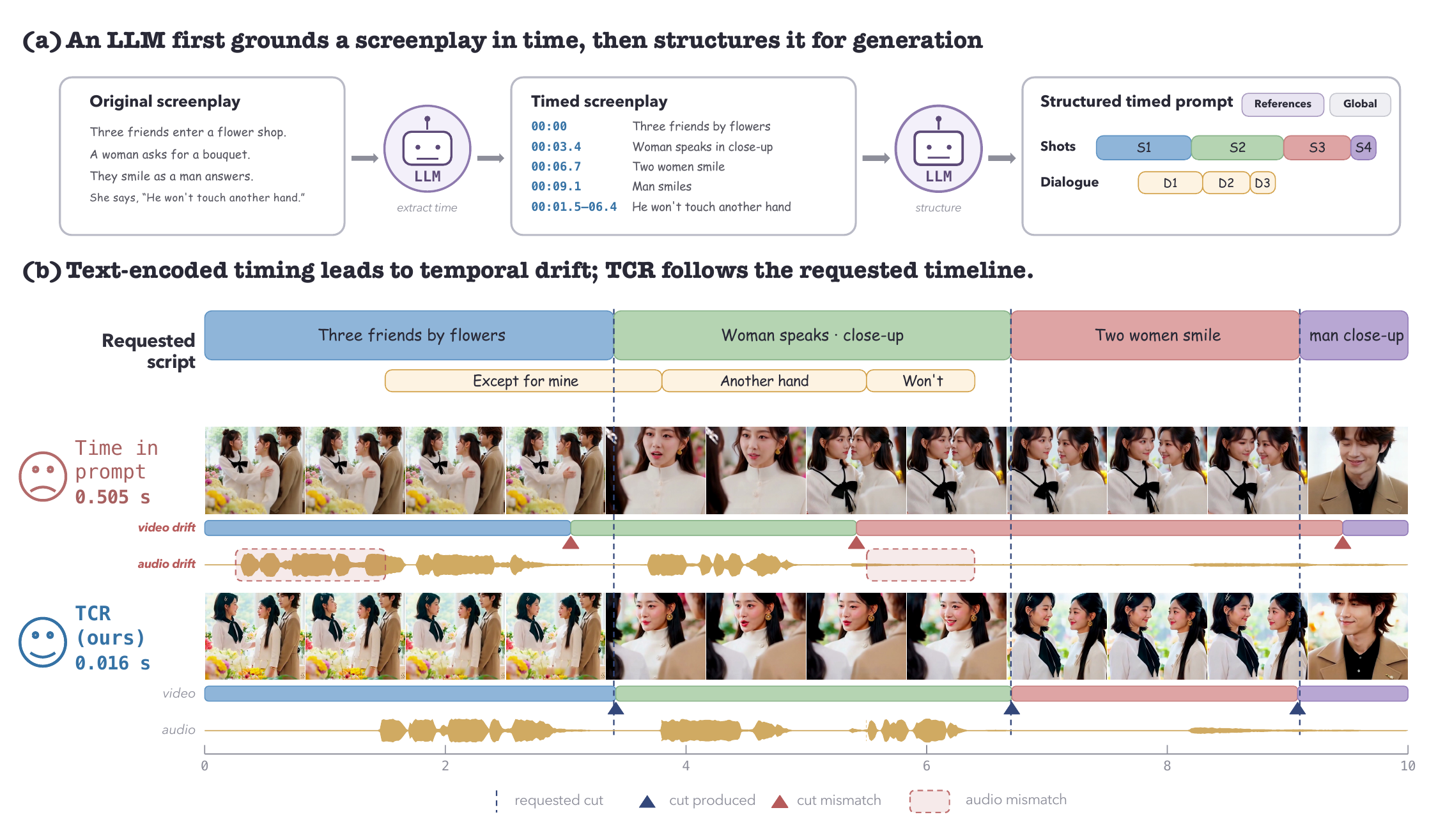}
  \caption{\textbf{(a)}~An LLM converts an original screenplay into a structured prompt that organizes references, global context, shots, and dialogue together with their precise timing. \textbf{(b)}~When script timing is encoded only in text, the generated cuts and speech fail to follow the requested timeline (red), whereas TCR follows the requested shot and dialogue timing more closely. Shot Boundary MAE is measured in seconds; lower is better.}
  \label{fig:teaser}
\end{figure*}

Joint audio-video generation has advanced rapidly, with recent models capable of synthesizing high-quality visual and acoustic content jointly from text prompts~\citep{ruan2023mmdiffusion,kondratyuk2024videopoet,liu2026javisdit,hacohen2026ltx2,low2025ovi,echo2026joyai}. As these capabilities mature, generative video is moving beyond isolated clip synthesis toward structured content-production workflows. One emerging direction is script-driven generation, which supports narrative video creation for applications such as short-form drama and advertising~\citep{tencent2026scriptavideo,zhou2026autocut}. Unlike conventional generation from a single text prompt~\citep{wan2025}, script-driven generation represents a scene as a structured prompt comprising multiple shot descriptions and dialogue lines. As illustrated in Figure~\ref{fig:teaser}(a), an original screenplay is first converted into this structured representation, which then guides the joint generation of video and audio.

Script-driven generation differs from conventional text-conditioned generation not only in how its input is organized, but also in the temporal control needed to realize that structure in the generated output. Each shot or dialogue prompt must guide both modalities during its designated time span~\citep{tencent2026scriptavideo}. Current joint generators align video and audio representations on a shared temporal axis~\citep{wang2024avdit,liu2026javisdit,hacohen2026ltx2}, yet encode script-specified shot and dialogue timing only in text, without explicitly associating each prompt with the corresponding temporal positions in either modality. Although video-focused methods have explored when local text prompts should influence video generation~\citep{yan2025presto,wu2025mindtime,shu2026tie}, they do not address how a structured script should jointly control video and audio. Consequently, the two modalities may remain synchronized with each other while jointly deviating from the script timeline, with shot transitions and speech occurring at the wrong times, as illustrated in Figure~\ref{fig:teaser}(b). Our central insight is therefore to extend temporal alignment beyond video and audio to include the structured script, representing the timing of each prompt as an explicit control signal aligned with the temporal coordinates of both modalities.

Realizing this control poses three challenges. First, shot and dialogue prompts may occupy distinct, partially overlapping time spans and must therefore remain independently controllable. A dialogue prompt may, for example, remain active across a shot boundary. Consequently, the two prompt types cannot be forced to share a single temporal segmentation. Second, learning this control requires fine-grained annotations of both shot boundaries and dialogue spans, whereas the initial annotations provide only coarse timing. Third, improving temporal accuracy must preserve the visual quality and audio-visual synchronization of the underlying joint generator. Together, these challenges motivate two complementary components: TCR for independent temporal control while preserving visual quality and audio-visual synchronization, and a coarse-to-fine data construction pipeline for refined temporal supervision.

TCR realizes our central insight by mapping each prompt’s specified timing onto the shared temporal axis of video and audio generation and routing its guidance to the corresponding positions in both modalities. It computes a duration-normalized routing score over temporal positions for each prompt and adds it as a bias to the video-text and audio-text cross-attention logits during both training and inference. This per-prompt additive design enables independent control of overlapping shot and dialogue prompts without modifying the original text, query, or key representations, helping preserve the visual quality and audio-visual synchronization of the underlying generator. To provide the fine-grained supervision required by TCR, we further develop a coarse-to-fine data construction pipeline that builds multi-shot clips with dialogue spanning shot transitions. Gemini supplies semantic annotations and coarse timing under our predefined script schema, which are subsequently refined using detected visual cuts and word-level speech alignment. The final shot and dialogue annotations are rounded to a 0.1 s grid.

We evaluate TCR on a test set of 200 scripts in two complementary settings: an end-to-end comparison with existing open-source joint generators and a controlled comparison with alternative temporal operators implemented on the same backbone. In the end-to-end comparison, TCR achieves the lowest shot-boundary error and the highest dialogue-timing accuracy among all evaluated systems. Compared with the strongest open-source baseline, it reduces Shot Boundary MAE by 96\%, from 1.11 s to 0.042 s, and increases Dialogue Acc@0.5 s from 28.3\% to 84.1\%, while maintaining competitive visual quality and audio-visual synchronization. In the controlled comparison, where only the temporal operator is varied, TCR reduces Shot Boundary MAE by more than 60\% relative to each of the two temporal baselines and again achieves the highest Dialogue Acc@0.5 s. An ablation using coarse instead of refined timing annotations substantially degrades both shot and dialogue accuracy, confirming the importance of the data construction pipeline. A user study further shows that TCR is preferred on all five evaluated dimensions: shot timing, dialogue timing, script fidelity, audio-visual synchronization, and overall quality.

Our main contributions are summarized as follows:
\begin{itemize}
  \item We identify a temporal gap in script-driven generation: video and audio
  may remain mutually synchronized yet fail to follow the script timeline. Our
  key insight is to extend their temporal alignment to the structured script,
  making each prompt's specified timing an explicit control signal for both
  modalities.
  \item We introduce Temporal Context Routing (TCR), which maps script timing
  onto the shared video-audio temporal axis and independently routes each
  prompt's guidance. We also develop a coarse-to-fine pipeline that builds
  multi-shot clips, uses Gemini under our predefined schema for semantic
  annotations and coarse timing, and refines shot and dialogue timing on a
  $0.1$\,s grid.
  \item We evaluate TCR on 200 test scripts. Compared with the strongest
  open-source baseline, TCR reduces Shot Boundary MAE by 96\%, from 1.11\,s to
  0.042\,s, and raises Dialogue Acc@0.5s from 28.3\% to 84.1\%, while
  maintaining competitive visual quality and audio-visual synchronization. A
  user study further shows that TCR is preferred across all five evaluated
  dimensions.
\end{itemize}

\section{Related Work}
\label{sec:related-work}

\textbf{Joint audio-visual generation.} Joint audio-visual generators build on diffusion models~\citep{ho2020ddpm}, diffusion transformers~\citep{peebles2023dit}, and flow matching~\citep{lipman2023flow}, with applications spanning multimodal and controllable video generation~\citep{zhou2026omnishow,searchgen,rationalrewards}. MM-Diffusion couples modality-specific denoisers through cross-modal attention~\citep{ruan2023mmdiffusion}, while later approaches adapt pretrained models or align cross-modal features~\citep{ishii2025simple,hajiali2025avlink}. AV-DiT shares a lightly adapted DiT backbone across modalities~\citep{wang2024avdit}. JavisDiT and JavisDiT++ introduce hierarchical spatio-temporal priors and unified optimization, respectively~\citep{liu2026javisdit,liu2026javisditpp}; Harmony combines cross-task training with synchronization-aware guidance~\citep{hu2026harmony}; and UniAVGen uses asymmetric, temporally aligned interactions~\citep{zhang2026uniavgen}. Other work explores synchronized conditional generation, native alignment, and asynchronous streams~\citep{song2026syncphony,wang2026hear,yariv2024diverse,ji2026native,li2026hallolive}. VideoPoet unifies multimodal tasks autoregressively~\citep{kondratyuk2024videopoet}, whereas LTX-2 uses interacting modality streams~\citep{hacohen2026ltx2}. These methods align audio and video with each other. Our work additionally aligns both modalities with the timing specified by a structured script, allowing each shot and dialogue prompt to control its designated time span.

\textbf{Structured and local prompting.} Beyond a monolithic prompt, methods use local or structured video conditions. Presto associates latent segments with subcaptions through segmented cross-attention~\citep{yan2025presto}, while ShotAdapter uses transition tokens and local attention masks for shot-specific control~\citep{kara2025shotadapter}. Specialized multi-shot generators model cinematic structure in different ways: HoloCine targets holistic long-form narratives, MultiShotMaster combines shot-aware rotary encodings with automatic annotation, and CineTrans learns cinematic transitions through masked diffusion~\citep{meng2026holocine,wang2026multishotmaster,wu2025cinetrans}. KeyVID and Audio-Sync Video Generation provide keyframe-aware and multi-stream temporal control for audio-synchronized visual generation~\citep{wang2025keyvid,weng2025audiosync}. MTSS factorizes an audio-visual description into grounded Reference, Shot, Event, and Global streams~\citep{tencent2026scriptavideo}. MTSS reconnects these streams through explicit identity and temporal links across the script. Video-focused methods ground local prompts only in the visual stream, whereas MTSS provides explicit temporal links without a mechanism that routes prompt timing through both video and audio conditioning pathways. Building on the MTSS schema, we align each prompt's specified timing with the temporal coordinates of both video and audio, allowing shot and dialogue prompts to remain independently controllable.

\textbf{Approaches to temporal control.} Temporal conditioning methods differ in where timing enters the pathway. \textbf{Access-based methods} use masks to expose prompt tokens only within designated temporal regions~\citep{yan2025presto,kara2025shotadapter}. \textbf{Representation-based methods} encode temporal structure in query--key interactions through RoPE variants~\citep{su2021roformer,wu2025mindtime,wang2026multishotmaster,shu2026tie}. For joint audio-visual synchronization, Cross-Modal Context Learning combines aligned RoPE with dynamic context routing~\citep{ma2026crossmodal}. Related approaches align modality streams through unified modeling, joint denoising, or synchronization features~\citep{liu2026javisditpp,wu2025doeshearinghelpseeing,song2026syncdit}, without aligning prompt-specific script timing with both streams. \textbf{Score-based methods} steer attention, latent states, queries, or logits toward target regions, often at inference time~\citep{cai2025ditctrl,schiber2026tempocontrol,xu2026switchcraft,zhang2026tsattn,chen2026promptrelay}. Related multimodal methods add conditioning for joint audio-video or video-to-audio generation~\citep{li2026mmcontrol,yang2026controlfoley}. Gaussian logit priors have modeled attention locality~\citep{yang2018localness,guo2019gaussian,kim2023gaussian}. TCR instead computes prompt-specific, duration-normalized routing scores from the time spans and applies them to video-text and audio-text cross-attention during training and inference, allowing overlapping shot and dialogue prompts to control both modalities independently.

\section{Method}
\label{sec:method}

\begin{figure*}[t]
  \centering
  \includegraphics[width=\textwidth]{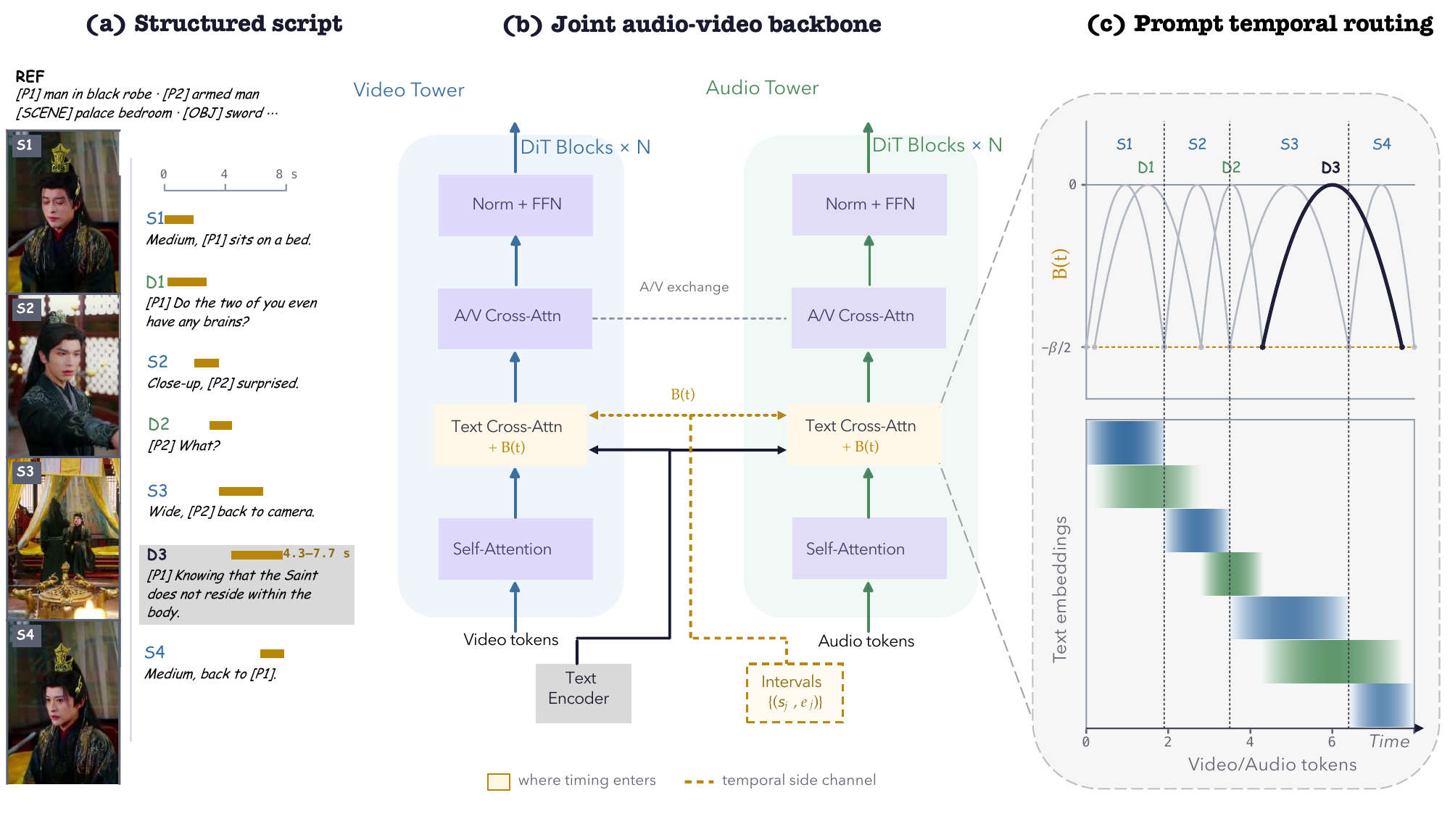}
  \caption{\textbf{Overview of Temporal Context Routing (TCR).}
  \textbf{(a)}~A structured script organizes reference, Shot, and dialogue
  prompts together with their timing. \textbf{(b)}~The shared text encoding
  conditions the video and audio towers, while prompt timing bypasses the text
  encoder and is supplied separately as $B(t)$ to both text cross-attention
  modules. \textbf{(c)}~Prompt temporal routing converts each prompt's timing
  into a duration-normalized routing profile that peaks at its center and
  reaches $-\beta/2$ at its boundaries. The resulting profiles route Shot
  (blue) and dialogue (green) guidance along the video-audio timeline.}
  \label{fig:method}
\end{figure*}

Given a structured script, our goal is to align the timing assigned to each shot
and dialogue prompt with the temporal coordinates used for video and audio
generation. As illustrated in Figure~\ref{fig:method}, each prompt's timing is
represented separately from its text encoding, and Temporal Context Routing
(\method) converts this timing into a duration-normalized routing score that is
added to the video--text and audio--text cross-attention logits. This per-prompt
construction allows shot and dialogue prompts to guide both modalities
according to their own timing. We first formalize the task and structured
script representation, and then describe the routing mechanism and the
coarse-to-fine data construction pipeline used to obtain temporal supervision.

\subsection{Task Formulation and Backbone}
\label{sec:task}

Let a structured script $S$ specify the content and timing of each shot and
dialogue prompt for a clip of duration $T$. After tokenization, the $j$th text
token inherits its parent prompt's interval
$I_j=[s_j,e_j]\subseteq[0,T]$. For modality $m\in\{v,a\}$, let
$\mathbf{q}_i^m$ denote a latent query at temporal coordinate $t_i^m$. Our goal
is to generate synchronized video $x^v$ and audio $x^a$ that follow both the
content and timing of $S$.

We build \method on LTX-2.3, a 22B-parameter joint audio-video generator whose
video and audio towers exchange information through audio--video
cross-attention and are conditioned on the shared script through separate text
cross-attention modules. \method modifies only the video--text and audio--text
cross-attention modules, aligning script timing with the temporal coordinates
of both modalities while leaving modality-specific self-attention and
audio--video cross-attention unchanged.

\subsection{Structured Script Representation}
\label{sec:mtss}

We condition the model on a structured script adapted from
MTSS~\citep{tencent2026scriptavideo}. It comprises four prompt types:
\textbf{Reference} identifies recurring people, scenes, and objects;
\textbf{Shot} describes shot content and camera attributes;
\textbf{Event} describes temporally localized audio events, with our
experiments focusing on spoken dialogue; and \textbf{Global} provides
clip-wide context. We retain only fields used to condition generation and omit
the Subtitle stream so that transcriptions of burned-in text do not condition
the model. Each prompt is assigned timing for routing: Shot and dialogue Event
prompts use their script-specified intervals, Global covers $[0,T]$, and
Reference timing is compiled from the shots in which the corresponding entity
appears. The complete schema and compilation rules are provided in
Appendix~\ref{app:prompts}.

\myparagraph{Timing extraction and token alignment.}
During serialization, we extract each \texttt{time\_range} field from the
structured script and omit it from the textual input. We then tokenize the
remaining script and use character-to-token offsets to compile the extracted
timing into a token-level map, assigning each token the interval of its parent
prompt. \method consumes this map alongside the shared text representation to
compute its routing scores. Appendix~\ref{app:prompts} details the handling of
repeated identifiers and tokens not associated with a specific prompt.

\subsection{Temporal Context Routing}
\label{sec:tcr}

For the interval $I_j=[s_j,e_j]$ assigned to token $j$, we define its center
and radius as
\begin{equation}
  c_j=\frac{s_j+e_j}{2},
  \qquad
  r_j=\max\left(\frac{e_j-s_j}{2},\epsilon\right),
\end{equation}
where $\epsilon=10^{-4}$ seconds handles degenerate intervals. For a modality
$m\in\{v,a\}$ and a latent query at temporal coordinate $t_i^m$, \method
defines the routing score
\begin{equation}
  B^{m}_{ij}
  =
  -\beta\frac{(t^{m}_i-c_j)^2}{2r_j^2}.
  \label{eq:tcr}
\end{equation}
Tokens without an associated prompt interval receive $B^{m}_{ij}=0$.

Let $\mathbf{h}_j$ denote the shared text representation of token $j$. The
text cross-attention module for modality $m$ projects it to
$\mathbf{k}_j^m=\mathbf{W}_K^m\mathbf{h}_j$. We then modify the
cross-attention logit as
\begin{equation}
  L^{m}_{ij}
  =
  \underbrace{
  \frac{(\mathbf{q}^{m}_i)^\top\mathbf{k}^{m}_j}{\sqrt{d_m}}
  }_{\mathclap{\text{\scriptsize semantic score}}}
  +
  \underbrace{
  B^{m}_{ij}
  }_{\mathclap{\text{\scriptsize routing score}}}
  +
  M_j,
  \label{eq:attention}
\end{equation}
where $d_m$ is the attention-head dimension and $M_j$ is the additive padding
mask, with $M_j=0$ for unmasked text tokens and $M_j=-\infty$ for padding
tokens.

\myparagraph{Additive temporal routing.}
For an unmasked text token, let
$S^{m}_{ij}=(\mathbf{q}^{m}_i)^\top\mathbf{k}^{m}_j/\sqrt{d_m}$. Its
unnormalized attention weight factorizes as
\begin{equation}
  \exp(S^{m}_{ij}+B^{m}_{ij})
  =
  \exp(S^{m}_{ij})\exp(B^{m}_{ij}).
  \label{eq:factorization}
\end{equation}
Thus, \method multiplicatively reweights the unnormalized attention induced by
the semantic score without modifying the text, query, or key representations.
For a nondegenerate interval, the routing score is $0$ at its center,
$-\beta/2$ at either endpoint, and decreases smoothly with normalized temporal
distance. Normalization by $r_j$ gives intervals of different durations the
same relative routing profile. We use $\beta=5$ throughout, yielding an
endpoint score of $-2.5$; Appendix~\ref{app:beta} discusses this choice.

\myparagraph{Independent routing across modalities.}
We evaluate Equation~\ref{eq:tcr} separately at the video and audio temporal
coordinates. Although the two modalities use different latent grids, both are
expressed in seconds relative to the same clip timeline. Computing a separate
routing score for every prompt and temporal position allows each shot or
dialogue prompt to retain its assigned timing independently of the boundaries
of other prompts. We apply \method during both LoRA adaptation and inference.
Because \method introduces no learnable parameters, we optimize only the LoRA
adapters~\citep{hu2022lora} under the original joint flow-matching objective.

\subsection{Coarse-to-Fine Data Construction}
\label{sec:data}

\begin{figure}[t]
  \centering
  \includegraphics[width=\linewidth]{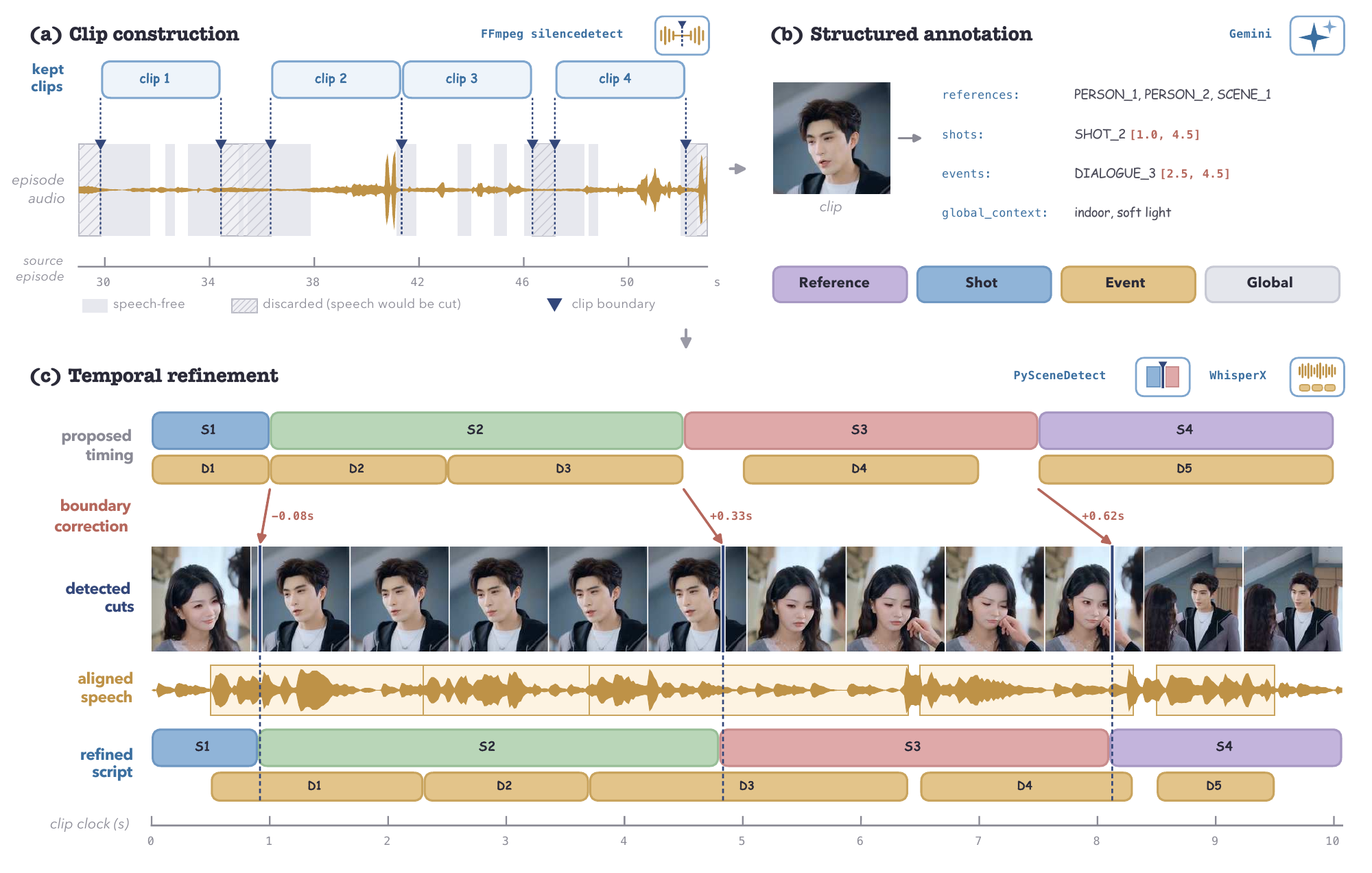}
  \caption{\textbf{Coarse-to-fine data construction.}
  \textbf{(a)}~Clip boundaries are selected within speech-free regions while
  internal shot transitions are preserved. \textbf{(b)}~Gemini converts each
  clip into Reference, Shot, Event, and Global prompts with coarse shot and
  dialogue timing. \textbf{(c)}~PySceneDetect corrects shot boundaries and
  WhisperX aligns dialogue to speech, producing refined temporal supervision
  on a $0.1\,\mathrm{s}$ grid.}
  \label{fig:data-pipeline}
\end{figure}

Fine-grained temporal control requires multi-shot training clips with accurate
shot and dialogue timing. We construct such examples in three stages, as
illustrated in Figure~\ref{fig:data-pipeline}.

\myparagraph{Clip construction.}
We combine speech-band silence detection with visual shot segmentation to
place clip boundaries within speech-free regions while preserving shot
transitions inside each clip. The resulting clips contain multiple shots and
retain dialogue that spans shot transitions, providing the temporal structure
needed for script-driven generation.

\myparagraph{Coarse script annotation.}
Gemini~\citep{gemini2026pro31} annotates each clip according to our predefined
script schema, generating Reference, Shot, Event, and Global prompts together
with coarse timestamps for Shot prompts and dialogue Events. Reference
identifiers avoid repeated appearance descriptions, and dialogue is preserved
in its original spoken language. We remove fields not used for conditioning,
including transcriptions of burned-in subtitles.

\myparagraph{Temporal refinement.}
We retain only scripts whose number of annotated Shot prompts matches that
inferred by an independently applied shot detector. For each
retained script, the detected cuts replace the coarse shot boundaries in
temporal order, preserving their correspondence with the semantic descriptions
of the Shot prompts. Thus, Gemini provides the script structure, while the
detector supplies localized visual boundaries.

For dialogue, WhisperX~\citep{bain2023whisperx} provides word-level speech
timestamps. We match the annotated dialogue lines to the transcription in
their original order and update each matched Event with the aligned transcript
and its start and end times. Unmatched or empty Events are removed without
discarding the rest of the script. Adjacent dialogue boundaries are adjusted
to resolve timing conflicts without constraining them to shot boundaries.
Finally, all refined shot and dialogue timestamps are rounded to a
$0.1\,\mathrm{s}$ grid.

\section{Experiments}
\label{sec:experiments}

\subsection{Experimental Setup}

\textbf{Implementation details.}
Our coarse-to-fine pipeline yields 57,022 training examples from two
short-drama collections. We evaluate on 200 test scripts containing 640 Shot
prompts and 441 dialogue prompts. No test script shares a source-media
identifier or caption hash with the training set. All models receive the same
shot and dialogue descriptions and target timing without first-frame
conditioning. Each model generates one output at $704\!\times\!1280$
resolution and $24\,\mathrm{fps}$ for the requested duration. Metrics are
averaged over the 200 outputs from each model.

\textbf{Baselines.}
We compare \method with Wan2.2~\citep{wan2025},
OVI~\citep{low2025ovi}, JoyAI-Echo~\citep{echo2026joyai}, and
LTX-2.3~\citep{hacohen2026ltx2}. Wan2.2 generates video only, whereas the
remaining models jointly generate video and audio. The baselines serialize
shot and dialogue timing as part of the script text. For \method, these timing
fields are removed before text encoding and supplied separately through the
timing map in Section~\ref{sec:mtss}. We additionally compare against
specialized multi-shot video generators using deterministic, model-specific
prompt adapters; the protocol and results are reported in
Appendix~\ref{app:multishot}.

\textbf{Metrics.}
Visual quality is measured using the Imaging Quality (IQ) and Aesthetic
Quality (AES) dimensions of VBench~\citep{huang2024vbench}. Temporal accuracy
is measured using Shot Boundary MAE, Shot IoU, exact shot-count accuracy, and
Dialogue Acc@0.5s. We further report WER using
Whisper-large-v3~\citep{radford2022whisper} and audio-visual synchronization
using SyncNet Sync-C and offset accuracy~\citep{chung2016out}. Metric
definitions and matching procedures are provided in
Appendix~\ref{app:metrics}.

\subsection{Main Results}
\label{sec:main-comparison}

\begin{table}[t]
  \centering
  \caption{End-to-end comparison on 200 test scripts.
  Table~\ref{tab:ablation-merged} reports controlled comparisons using the
  same backbone and training setup.}
  \vspace{2mm}
  \renewcommand{\arraystretch}{1.2}
  \label{tab:main-comparison}
  \small
  \setlength{\tabcolsep}{3.2pt}
  \begin{tabular}{@{}lccccccccc@{}}
    \toprule
    & \multicolumn{2}{c}{Video quality} & \multicolumn{4}{c}{Temporal accuracy} & \multicolumn{3}{c}{Speech \& AV sync} \\
    \cmidrule(lr){2-3}\cmidrule(lr){4-7}\cmidrule(lr){8-10}
    Method
      & IQ $\uparrow$ & AES $\uparrow$
      & \makecell{Shot\\B-MAE (s) $\downarrow$} & \makecell{Shot\\IoU $\uparrow$}
      & \makecell{Shot\\Count Acc.\\(\%) $\uparrow$} & \makecell{Acc@0.5s\\(\%) $\uparrow$}
      & \makecell{WER\\(\%) $\downarrow$} & Sync-C $\uparrow$
      & \makecell{Off.\ Acc\\(\%) $\uparrow$} \\
    \midrule
    Wan2.2
      & \underline{0.6820} & 0.5150 & 2.31 & 0.422 & 9.0 & - & - & - & - \\
    OVI
      & 0.6640 & \textbf{0.5602} & 1.84 & 0.457 & 17.0 & 8.6 & 56.9 & 2.37 & 17.2 \\
    JoyAI-Echo
      & 0.6680 & 0.5167 & 1.81 & 0.464 & 15.5 & 23.2 & \underline{11.9} & \underline{2.71} & 7.3 \\
    LTX-2.3
      & 0.6819 & 0.5354 & \underline{1.11} & \underline{0.532} & \underline{36.0} & \underline{28.3} & 14.0 & 2.55 & \textbf{31.2} \\
    \midrule
    \rowcolor{abstitle!7}
    \method (ours)
      & \textbf{0.7032} & \underline{0.5477}
      & \textbf{0.042} & \textbf{0.957} & \textbf{93.0} & \textbf{84.1}
      & \textbf{8.48} & \textbf{2.78} & \underline{30.5} \\
    \bottomrule
  \end{tabular}
\end{table}

\textbf{Quantitative comparison.}
As shown in Table~\ref{tab:main-comparison}, \method achieves the most accurate
shot and dialogue timing among the evaluated models. Relative to the strongest
baseline on each temporal metric, it reduces Shot Boundary MAE by 96\%, from
$1.11\,\mathrm{s}$ to $0.042\,\mathrm{s}$, and increases Shot IoU from 0.532
to 0.957. Exact shot-count accuracy similarly rises from 36.0\% to 93.0\%,
while Dialogue Acc@0.5s increases from 28.3\% to 84.1\%. Beyond temporal
accuracy, \method achieves the highest IQ and Sync-C and the lowest WER, while
remaining competitive on the other quality and synchronization metrics.

\begin{figure}[t]
  \centering
  \includegraphics[width=\linewidth]{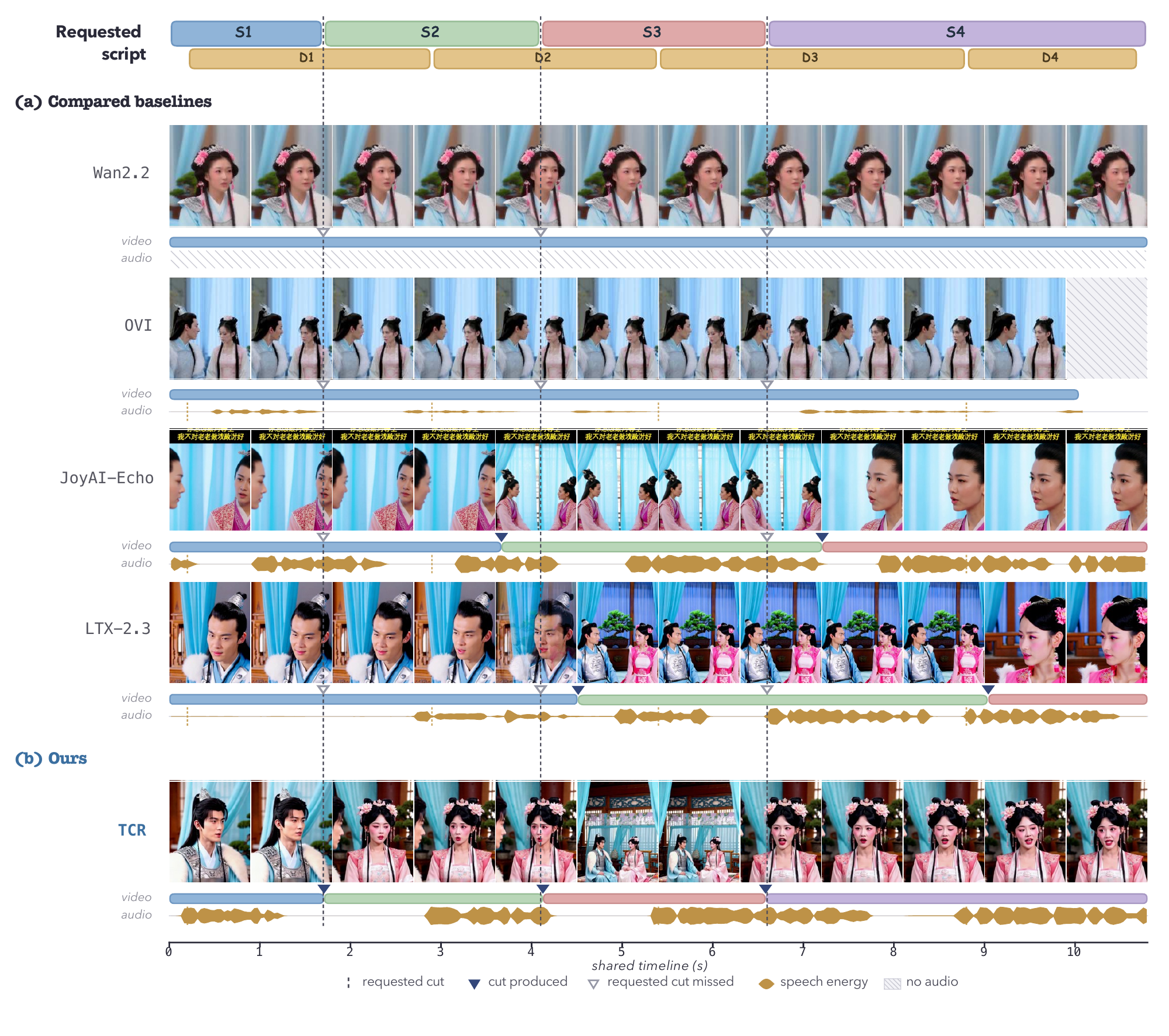}
  \caption{\textbf{Qualitative comparison on a four-shot script.}
  The requested Shot and dialogue timing is shown at the top.
  \textbf{(a)}~Each baseline misses or delays at least one of the three
  requested shot transitions; Wan2.2 generates no audio. \textbf{(b)}~\method
  produces all three cuts near their requested times and follows the specified
  four-shot structure. Speech-energy traces show dialogue activity on the same
  timeline.}
  \label{fig:system-qualitative}
\end{figure}

\textbf{Qualitative comparison.}
Figure~\ref{fig:system-qualitative} compares the generated outputs for a script
containing four Shot prompts and four dialogue Events. The video lanes display
generated frames and detected shot transitions, while the audio lanes place
speech-energy traces against the requested dialogue timing. \method produces
all three requested shot transitions close to their target times and follows
the specified four-shot structure. In contrast, each baseline misses or delays
at least one transition. For the audio-generating models, the speech-energy
traces further reveal discrepancies from the requested dialogue timing. The
visualization thus provides a direct view of the script-alignment errors
captured by the temporal metrics in Table~\ref{tab:main-comparison}.

\textbf{Human evaluation.}
We further conduct a blinded pairwise study comparing \method with LTX-2.3 and
JoyAI-Echo on 16 randomly sampled cases, with eight cases per comparator.
Twenty-eight participants evaluate shot timing,
dialogue timing, script fidelity, audio-visual synchronization, and overall
preference, with ties explicitly allowed. The complete protocol and
interface are provided in Appendix~\ref{app:user-study}. As shown in
Figure~\ref{fig:user-study}, \method is preferred over both comparators across
all five dimensions. With ties included in the denominator, \method receives
72.3\% of the overall-preference votes against LTX-2.3 and 83.9\% against
JoyAI-Echo. These results indicate that the gains in temporal accuracy are also
reflected in perceived script execution.

\begin{figure}[t]
  \centering
  \includegraphics[width=\textwidth]{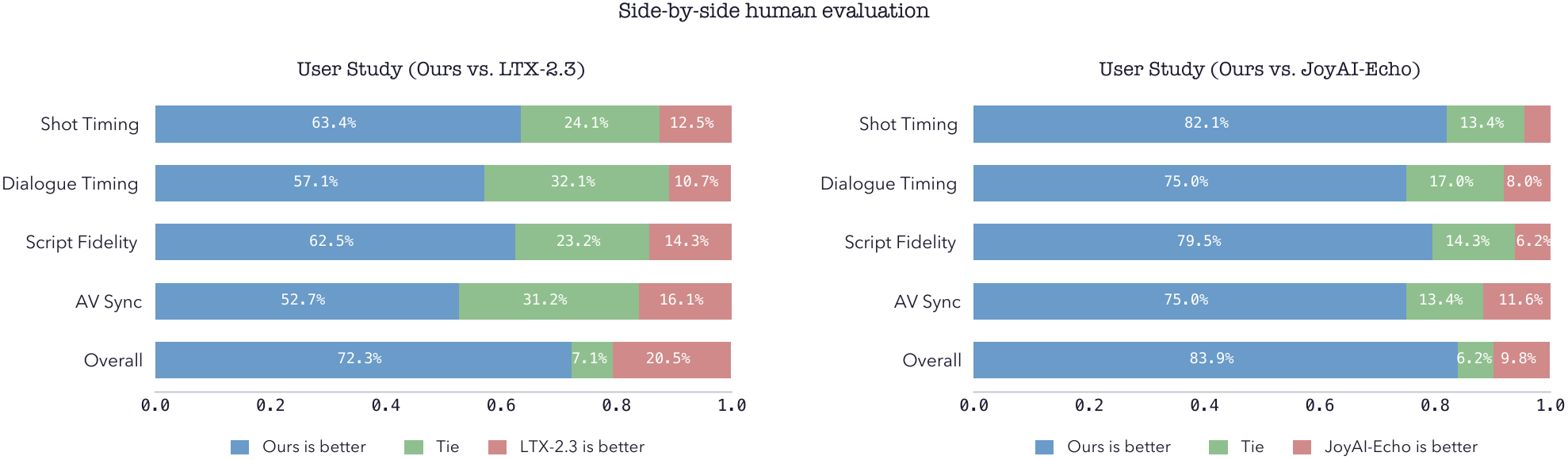}
  \caption{\textbf{Pairwise human evaluation.} Tie-inclusive vote shares for
  \method against LTX-2.3 (left) and JoyAI-Echo (right) across five dimensions,
  based on 16 randomly sampled cases and 28 participants.}
  \label{fig:user-study}
\end{figure}

\subsection{Ablation Studies and Analysis}
\label{sec:temporal-results}

To isolate the effect of the temporal operator, we compare \method with matched
joint audio-video implementations of Gaussian Interval RoPE and a hard interval
mask, using the same backbone, timing map, refined training data, and
optimization settings. We include \emph{Intervals as text}, which retains
timing in the serialized script without an attention-level operator, as an
auxiliary reference rather than a matched operator. We separately ablate
temporal refinement and prompt assignment across the video and audio branches.
Table~\ref{tab:ablation-merged} summarizes the results. Operator and training
details are provided in Appendices~\ref{app:operators} and~\ref{app:impl}.

\begin{table}[t]
  \centering
  \caption{Controlled comparison of temporal operators and TCR ablations.
  All rows except \emph{Intervals as text} use the shared training recipe.
  Bold/underline: best/second among these matched rows.}
  \vspace{2mm}
  \renewcommand{\arraystretch}{1.2}
  \label{tab:ablation-merged}
  \small
  \setlength{\tabcolsep}{1.3pt}
  \begin{tabular}{@{}lccccccccc@{}}
    \toprule
    & \multicolumn{2}{c}{Video quality} & \multicolumn{4}{c}{Temporal accuracy} & \multicolumn{3}{c}{Speech \& AV sync} \\
    \cmidrule(lr){2-3}\cmidrule(lr){4-7}\cmidrule(lr){8-10}
    Method
      & IQ $\uparrow$ & AES $\uparrow$
      & \makecell{Shot\\B-MAE (s) $\downarrow$} & \makecell{Shot\\IoU $\uparrow$}
      & \makecell{Shot\\Count Acc.\\(\%) $\uparrow$} & \makecell{Acc@0.5s\\(\%) $\uparrow$}
      & \makecell{WER\\(\%) $\downarrow$} & Sync-C $\uparrow$
      & \makecell{Off.\ Acc\\(\%) $\uparrow$} \\
    \midrule
    \multicolumn{10}{@{}l}{\itshape Auxiliary reference} \\
    Intervals as text
      & 0.7010 & 0.5548 & 0.601 & 0.629 & 83.0 & 43.8
      & 9.38 & 2.86 & 33.7 \\
    \midrule
    \multicolumn{10}{@{}l}{\itshape Matched temporal operators} \\
    Gaussian Interval RoPE
      & 0.6931 & 0.5450 & 0.113 & 0.915 & \underline{92.0} & 82.8
      & 10.64 & 2.67 & 24.5 \\
    \makecell[l]{Hard interval mask}
      & 0.7005 & \underline{0.5510} & 0.108 & 0.902 & \underline{92.0} & \underline{83.7}
      & 9.32 & \underline{2.71} & 26.0 \\
    \rowcolor{violet!7}
    \method
      & 0.7032 & 0.5477 & \textbf{0.042} & \textbf{0.957} & \textbf{93.0} & \textbf{84.1}
      & \textbf{8.48} & \textbf{2.78} & \textbf{30.5} \\
    \midrule
    \multicolumn{10}{@{}l}{\itshape TCR ablations} \\
    \makecell[l]{w/ separate A/V prompts}
      & \underline{0.7067} & \textbf{0.5548} & \underline{0.047} & \underline{0.953} & 91.5 & 82.1
      & \underline{8.86} & \underline{2.71} & 29.3 \\
    \makecell[l]{ w/o refinement}
      & \textbf{0.7086} & 0.5460 & 0.375 & 0.715 & 83.0 & 37.6
      & 9.22 & 2.67 & \underline{30.4} \\
    \bottomrule
  \end{tabular}
\end{table}

\textbf{Temporal operators.}
Among the three matched temporal operators, \method performs best on every
shot-timing metric. It reduces Shot Boundary MAE by more than 60\% relative to
both Gaussian Interval RoPE and the hard mask, reaching
$0.042\,\mathrm{s}$, while increasing Shot IoU to 0.957 and exact shot-count
accuracy to 93.0\%. \method also achieves the highest Dialogue Acc@0.5s at
84.1\%, compared with 83.7\% for the hard mask and 82.8\% for Gaussian
Interval RoPE. Checkpoint-level results are provided in
Appendix~\ref{app:visual-curves}.

\textbf{Temporal supervision.}
Training with coarse rather than refined annotations substantially degrades
temporal control. Shot Boundary MAE increases from $0.042\,\mathrm{s}$ to
$0.375\,\mathrm{s}$, while Dialogue Acc@0.5s decreases from 84.1\% to 37.6\%.
IQ and AES remain comparable, indicating that the degradation primarily
affects temporal accuracy. These results demonstrate the importance of the
fine-grained supervision produced by our data construction pipeline.

\textbf{Prompt sharing.}
The \emph{Separate A/V prompts} variant provides Shot prompts only to the video
branch and dialogue prompts only to the audio branch. Although its shot timing
remains close to that of \method, its Dialogue Acc@0.5s, WER, Sync-C, and
offset accuracy all degrade. Providing both prompt types to both branches
therefore better supports joint temporal control and audio-visual coordination.

\textbf{Quality and synchronization.}
IQ and AES vary by less than 2.5\% in relative terms across the matched
operators and ablations. Among the matched temporal operators, \method also
achieves the lowest WER and the highest Sync-C and offset accuracy. Its gains
in temporal control therefore do not come at the expense of visual quality or
audio-visual synchronization.

\section{Conclusion}
\label{sec:conclusion}

This work addresses the missing temporal alignment between structured scripts
and joint audio-video generation. Although existing generators align video and
audio on a shared temporal axis, the shot and dialogue timing explicitly
specified by a script is represented only implicitly in text conditioning.
Consequently, the two modalities may remain synchronized with each other while
jointly deviating from the script timeline. We introduce Temporal Context
Routing (TCR), which maps script-specified timing onto the shared video-audio
temporal axis and routes each prompt's guidance to the corresponding positions
in both modalities. We further develop a coarse-to-fine data construction
pipeline that provides accurate shot and dialogue timing for temporal
supervision. 
On 200 test scripts, TCR reduces Shot Boundary MAE by 96\%, from
$1.11\,\mathrm{s}$ to $0.042\,\mathrm{s}$, and raises Dialogue Acc@0.5s from
28.3\% to 84.1\% compared with the strongest baseline, while maintaining
competitive visual quality and audio-visual synchronization. Controlled
comparisons and ablations further demonstrate the effectiveness of temporal
routing and the importance of refined supervision, while a user study shows
that TCR is preferred on all five evaluated dimensions. Overall, these results
show that extending temporal alignment to the structured script enables more
accurate script-driven generation without compromising the visual quality or
audio-visual synchronization of the underlying joint generator.

\bibliographystyle{colm2024_conference}
\bibliography{main}

\appendix
\clearpage

\section{Temporal Operator Implementations}
\label{app:operators}

We compare \method with two matched temporal operators implemented on the same
joint audio-video backbone: a hard interval mask and Gaussian Interval RoPE.
All three use the same timing map, training data, LoRA configuration, and
optimization settings; they differ only in how timing enters text
cross-attention. \emph{Intervals as text} is retained as an auxiliary reference
and is not part of this matched comparison.

\paragraph{Intervals as text.}
Numeric \texttt{time\_range} fields remain in the serialized script and are
processed by the text encoder together with the prompt content. No separate
attention-level temporal operator is applied.

\paragraph{Hard interval mask.}
This operator replaces Equation~\ref{eq:tcr} with
\begin{equation}
  B^{m,\mathrm{hard}}_{ij}=
  \begin{cases}
    0, & t^{m}_i\in[s_j,e_j]\ \text{or token $j$ is sentinel},\\
    -\infty, & \text{otherwise}.
  \end{cases}
  \label{eq:hard-mask}
\end{equation}
It exposes a prompt to queries inside its assigned interval and masks it at all
other temporal positions.

\paragraph{Gaussian Interval RoPE.}
We implement an interval-aware rotary baseline based on
RoPE~\citep{su2021roformer} and temporal rotary encodings~\citep{wu2025mindtime,shu2026tie}.
A query is rotated at its temporal coordinate and a text key at the center of
its assigned interval. After mapping seconds to the model's normalized time
axis, let $\widehat{r}_j$ denote the interval radius. At frequency
$\omega_\ell$, the key channel is scaled by
\begin{equation}
  g_{j\ell}=\frac{\exp[-\tfrac{1}{2}(\alpha\omega_\ell \widehat{r}_j)^2]}
  {\operatorname{mean}_{\ell'}\exp[-\tfrac{1}{2}(\alpha\omega_{\ell'}\widehat{r}_j)^2]},
  \label{eq:gaussian-rote}
\end{equation}
a Gaussian kernel that accounts for interval duration. Unlike \method, this
operator incorporates content and timing jointly in the query-key geometry. We
use $\alpha=1$ and a frequency scale of $2.5\times10^{-3}$ for both modality
towers.

\section{Comparison with Multi-Shot Video Generators}
\label{app:multishot}

We further compare \method with three specialized multi-shot video generators:
CineTrans~\citep{wu2025cinetrans},
MultiShotMaster~\citep{wang2026multishotmaster}, and
HoloCine~\citep{meng2026holocine}. We evaluate every method on the same 200
test scripts, and all three baselines completed all 200 generations and
evaluations. Because these baselines generate silent video, this comparison is
restricted to shot timing and visual quality; the joint audio-video comparison
remains in Table~\ref{tab:main-comparison}.

\paragraph{Deterministic prompt adaptation.}
We adapt the structured scripts to each model's native input format without
LLM-based rewriting or recaptioning. Across all methods, the adapters preserve
the same characters, actions, scenes, shot scales, camera angles, and camera
motions, map entity identifiers to neutral subject indices, and exclude
dialogue, sound effects, and global-audio fields. Let $G$ be the global caption
containing entity and scene descriptions, the overall story, and visual style,
and let $S_i$ be the visual and camera description of shot $i$. CineTrans takes
$[S_1,\ldots,S_K,G]$ and receives duration-proportional shot lengths separately
through \texttt{mask\_info} on its latent timeline; its total length is rounded
up to a valid $4n+1$ frame count at 16\,fps. MultiShotMaster receives one prompt
$\text{``Story: }G\ \text{Now: }S_i\text{''}$ per shot. We quantize the requested
durations at 15\,fps to form its shot intervals, and remove the model's temporary
$4n+1$ padding before concatenating the shots. HoloCine uses
\texttt{[global caption]} $G$, followed by \texttt{[per shot caption]} $S_1$
and \texttt{[shot cut]} separators before subsequent shots. Its cut positions
are mapped from the requested temporal boundaries at 15\,fps. Under its
500-token limit, 193 scripts require no compression, four drop only the overall
story and visual-style fields, and three additionally truncate only $G$; all
per-shot descriptions remain intact.

\begin{table}[t]
  \centering
  \caption{Comparison with specialized multi-shot video generators on all 200
  test scripts. All baselines use the deterministic adapters described above.
  B-MAE denotes Shot Boundary MAE; IQ and AES are the Imaging and Aesthetic
  Quality dimensions of VBench.}
  \label{tab:multishot-comparison}
  \vspace{2mm}
  \renewcommand{\arraystretch}{1.15}
  \small
  \setlength{\tabcolsep}{4.2pt}
  \begin{tabular}{@{}lccccc@{}}
    \toprule
    Method & \makecell{B-MAE (s)\\$\downarrow$} & \makecell{Shot IoU\\$\uparrow$}
      & \makecell{Count Acc. (\%)\\$\uparrow$} & \makecell{IQ\\$\uparrow$}
      & \makecell{AES\\$\uparrow$} \\
    \midrule
    CineTrans & 0.278 & 0.849 & 62.0 & \textbf{0.7089} & \textbf{0.5629} \\
    MultiShotMaster & \underline{0.054} & \underline{0.945} & \underline{79.0}
      & 0.6696 & 0.5448 \\
    HoloCine & 0.139 & 0.859 & 75.0 & 0.6700 & 0.5432 \\
    \midrule
    \rowcolor{abstitle!7}
    \method (ours) & \textbf{0.042} & \textbf{0.957} & \textbf{93.0}
      & \underline{0.7032} & \underline{0.5477} \\
    \bottomrule
  \end{tabular}
\end{table}

As shown in Table~\ref{tab:multishot-comparison}, \method achieves the best
temporal accuracy on all three measures. Relative to the strongest multi-shot
baseline, MultiShotMaster, it reduces Shot Boundary MAE from 0.054\,s to
0.042\,s, increases Shot IoU from 0.945 to 0.957, and raises exact shot-count
accuracy from 79.0\% to 93.0\%. CineTrans obtains the highest VBench quality
scores, while \method ranks second on both IQ and AES and remains close in
absolute value. These results indicate that routing script time directly into
the conditioning pathway improves shot-level temporal control while preserving
competitive visual quality, even against systems designed specifically for
multi-shot video generation.

%
%
%
%
%

\section{Visual-Timing Optimization Trajectories}
\label{app:visual-curves}

\begin{figure*}[t]
  \centering
  \includegraphics[width=0.98\linewidth]{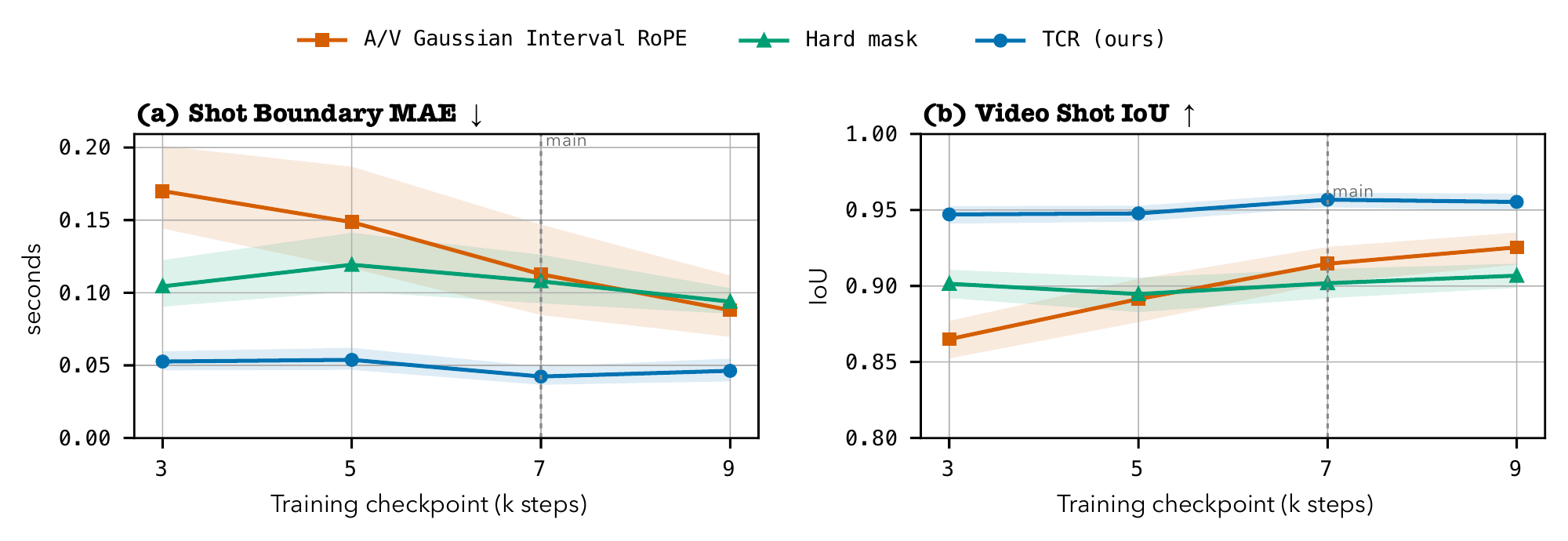}
  \caption{\textbf{Shot-timing performance across training checkpoints.}
  \textbf{(a)}~Shot Boundary MAE and \textbf{(b)}~Shot IoU on the same 200 test
  scripts. Shaded regions denote 95\% prompt-bootstrap intervals, and the
  dashed line marks the 7k checkpoint reported in the main tables. \method
  performs best on both metrics at every checkpoint.}
  \label{fig:learning-curves}
\end{figure*}

Figure~\ref{fig:learning-curves} tracks visual timing across training. \method
achieves the lowest Shot Boundary MAE and highest Shot IoU at every evaluated
checkpoint. The advantage is already present at 3,000 steps and remains stable
through 9,000 steps, while the main tables report the 7,000-step checkpoint.

\section{Qualitative Ablation Examples}
\label{app:qualitative}

\begin{figure}[t]
  \centering
  \includegraphics[width=0.95\linewidth]{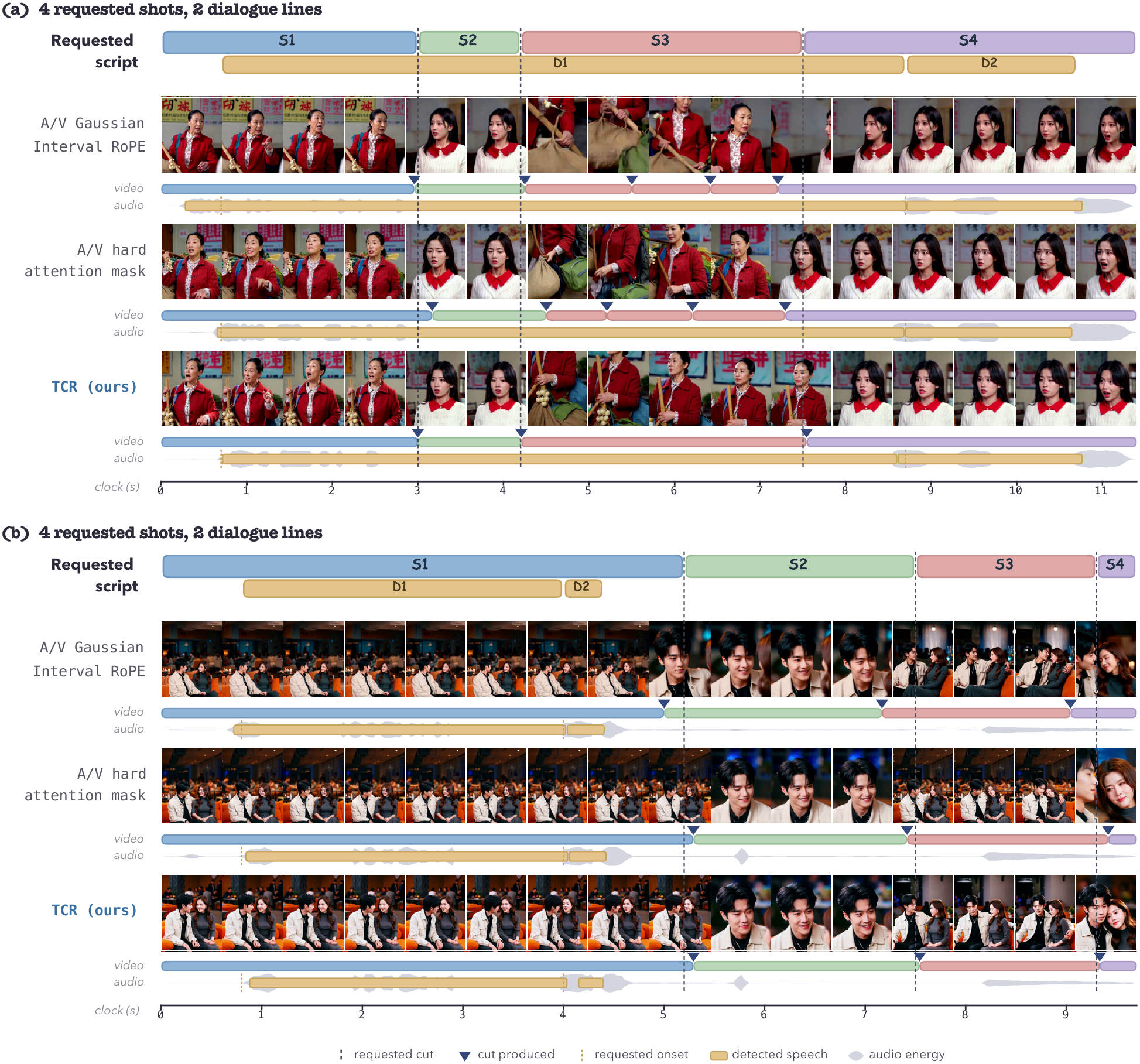}
  \caption{\textbf{Qualitative comparison of temporal operators.} Requested
  shot and dialogue timing is shown above each example. \textbf{(a)}~Gaussian
  Interval RoPE and the hard mask produce two extra cuts, whereas \method
  matches all four requested shots. \textbf{(b)}~Gaussian Interval RoPE places
  all three cuts early, while the hard mask and \method align them closely with
  the requested boundaries. Triangles mark produced cuts; the audio lanes show
  detected speech and energy.}
  \label{fig:qualitative}
\end{figure}

Figure~\ref{fig:qualitative} provides script-level qualitative comparisons
corresponding to the ablation results in Table~\ref{tab:ablation-merged}. All
three temporal operators are trained and evaluated under identical settings,
with their outputs displayed against the requested script timeline. The
examples visualize differences in shot placement and dialogue timing,
illustrating the temporal behaviors reflected in the aggregate metrics.

\section{Human Evaluation Protocol}
\label{app:user-study}

\begin{figure}[t]
  \centering
  \includegraphics[width=\linewidth]{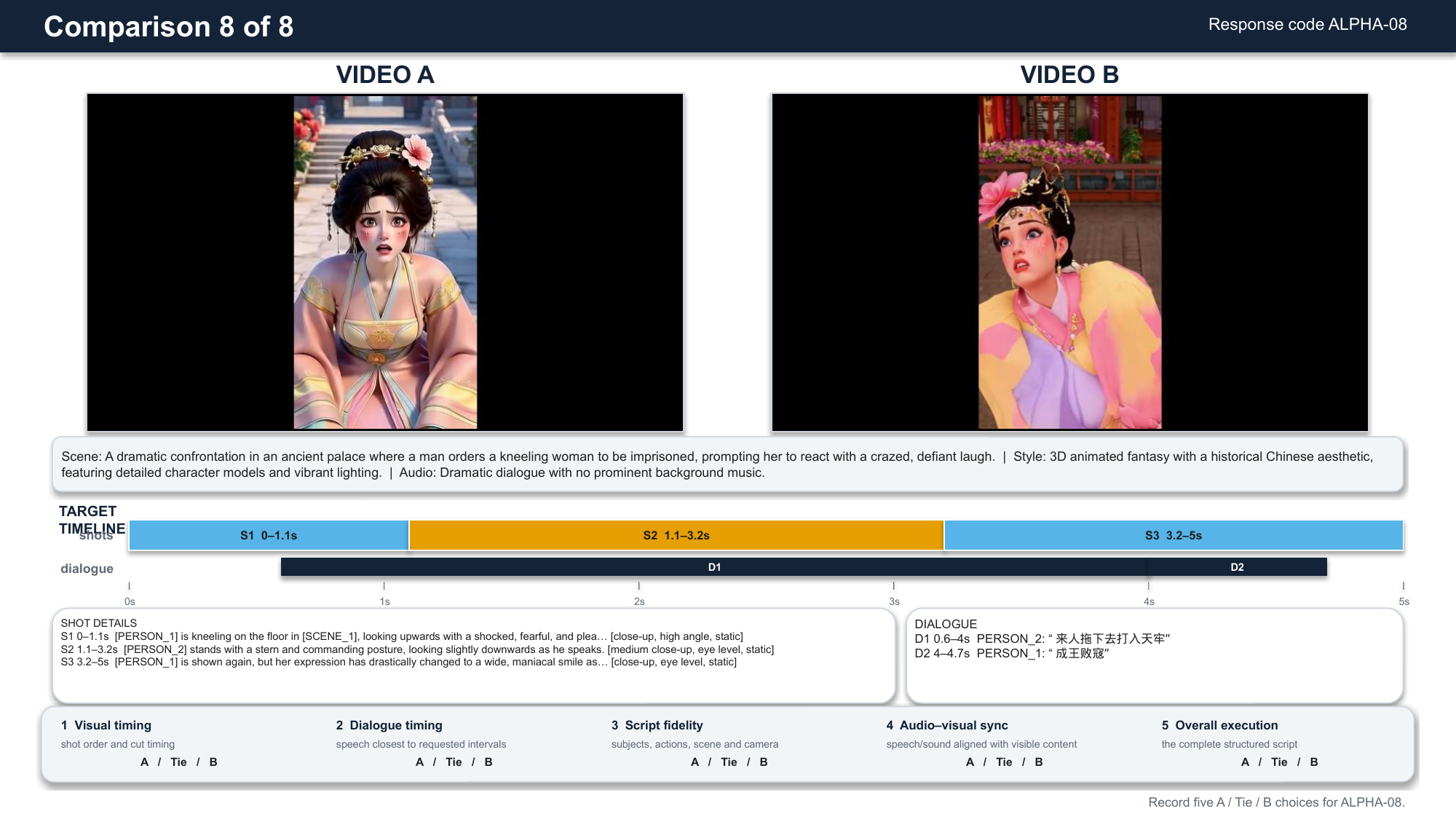}
  \caption{\textbf{Human-evaluation interface.} Videos are presented
  anonymously as A and B with balanced ordering. Raters view the target shot
  and dialogue timeline and make independent A/Tie/B choices for shot timing,
  dialogue timing, script fidelity, audio-visual synchronization, and overall
  execution.}
  \label{fig:user-study-interface}
\end{figure}

We randomly sample 16 test cases, with eight used for the comparison with
LTX-2.3 and eight for JoyAI-Echo. Twenty-eight participants complete all 16
trials. Each trial presents \method and one comparator anonymously as Video A
and Video B; side assignment is balanced across trials, and JoyAI-Echo outputs
are generated at $704\!\times\!1280$ so that all videos share the same portrait
canvas. Participants make independent A/Tie/B choices for shot timing,
dialogue timing, script fidelity, audio-visual synchronization, and overall
execution. Figure~\ref{fig:user-study-interface} shows the evaluation interface.

\section{Metric Details}
\label{app:metrics}

\myparagraph{Shot timing.}
For a requested shot $G=[g_s,g_e]$ and a detected shot
$P=[p_s,p_e]$, temporal intersection-over-union is
\begin{equation}
  \operatorname{IoU}(G,P)=
  \frac{\max(0,\min(g_e,p_e)-\max(g_s,p_s))}
       {\max(g_e,p_e)-\min(g_s,p_s)}.
\end{equation}
The corresponding boundary error is
\begin{equation}
  E_{\mathrm{bnd}}(G,P)=\frac{|g_s-p_s|+|g_e-p_e|}{2}.
\end{equation}
Requested and detected shots are paired in temporal order when their counts
agree and greedily by highest positive IoU otherwise. Shot Boundary MAE and
Shot IoU average over the resulting matched pairs. Matched-shot coverage and
exact shot-count accuracy separately capture missing and additional shots. All
test scripts pass the Gemini-detector shot-count consistency check, and the
same frozen detector is applied to every generated output. At 24\,fps, the
reported $0.042$\,s Shot Boundary MAE is approximately one output frame.

\myparagraph{Dialogue timing.}
WhisperX~\citep{bain2023whisperx} provides word-level speech timestamps, which
are matched monotonically to the requested dialogue lines. Detection rate is
the fraction of requested lines with a match. Start, end, and Boundary MAE are
computed over matched lines. Event IoU assigns zero to an unmatched dialogue,
and Acc@0.5s is the fraction of all requested dialogue prompts whose matched
onset and offset errors are both within $0.5$\,s.

\myparagraph{Speech fidelity and audio-visual synchronization.}
WER compares the requested dialogue with transcripts from
Whisper-large-v3~\citep{radford2022whisper}. Sync-C and offset accuracy are
computed with SyncNet~\citep{chung2016out} on speech-active within-shot
segments. Offset accuracy is the fraction of segments whose estimated
audio-video offset is within one frame at 24\,fps.

\myparagraph{Visual quality.}
We report the Imaging Quality and Aesthetic Quality dimensions of VBench.
Whole-clip consistency and motion-smoothness are not included because outputs
that omit requested cuts can score favorably on these dimensions despite
failing the target temporal structure.

\section{Choosing the Routing Strength \texorpdfstring{$\beta$}{beta}}
\label{app:beta}

\myparagraph{Endpoint-retention parameterization.}
Equation~\ref{eq:tcr} depends on the query time only through the normalized
temporal distance $u=(t^{m}_i-c_j)/r_j$, so the routing term contributes a
multiplicative factor
\begin{equation}
  g_\beta(u)=\exp\!\left(-\tfrac{1}{2}\beta u^{2}\right)
  \label{eq:beta-profile}
\end{equation}
to the unnormalized attention weight of Equation~\ref{eq:attention}. Because
an interval endpoint has $|u|=1$, $\beta$ can be interpreted through the
endpoint-retention ratio $\varepsilon=g_\beta(1)$: the fraction of the center
weight retained at either endpoint. Solving for $\beta$ gives
\begin{equation}
  \beta=2\ln(1/\varepsilon),
  \label{eq:beta-from-eps}
\end{equation}
We choose $\varepsilon=0.1$, corresponding to one order of magnitude of
attenuation from the interval center to either endpoint. This gives
$\beta=2\ln 10=4.605$, which we round to $\beta=5$. The resulting endpoint
retention is $e^{-5/2}=0.082$. We use this fixed value in every block, modality,
and attention head during both training and inference; it is neither learned
nor tuned per prompt.

\section{Implementation Details}
\label{app:impl}

\myparagraph{Temporal coordinates and batching.}
Video query times are the midpoints of the temporal cells represented by the
video latents. Audio query times are computed from audio patch positions using
the hop length and sample rate. Both are converted to seconds before applying
the temporal operator. We use $\epsilon=10^{-4}$\,s in
Equation~\ref{eq:tcr}, assign Global prompts to $[0,T]$, and apply no routing
score to sentinel tokens.

After tokenization, the timing map is repacked to match the text connector,
including left padding and special tokens. The compiled tensor
$\mathbf{I}\in\mathbb{R}^{N\times2}$ is padded to $B\times N\times2$. The
padding mask has shape $B\times1\times1\times N$, and the routing score
$B^m$ has shape $B\times1\times Q_m\times N$ and is shared across attention
heads. Classifier-free guidance follows the original LTX-2.3 conditional and
unconditional branch construction; \method adds no CFG-specific parameters.

\myparagraph{Training and inference.}
The three matched temporal operators use LoRA rank and scale 128, a learning
rate of $10^{-4}$, 500 warm-up steps, cosine decay, a first-frame conditioning
probability of 0.5 during training, and an audio loss weight of 1. The main
tables report the 7,000-step checkpoint; Figure~\ref{fig:learning-curves} also
evaluates 3,000, 5,000, and 9,000 steps. At inference, we use 30 sampling steps,
guidance 4.0, spatiotemporal guidance 1.0 at block 29, seed 42,
$704\times1280$ resolution, and 24 fps, without first-frame conditioning.

\section{Structured Script and Temporal Compilation}
\label{app:prompts}

The abbreviated record below shows the structured script information provided
to every evaluated model. Wan2.2, OVI, JoyAI-Echo, LTX-2.3, and the
\emph{Intervals as text} reference retain numeric \texttt{time\_range} fields
in the serialized prompt. For the three matched attention-level operators, the
same source record is passed through a compiler that removes these numeric
ranges before text encoding and supplies them separately through the timing
map. All variants therefore receive the same prompt content, requested timing,
and clip duration.

The compiler aligns the character span of each prompt with its token positions
and assigns each token the timing of its parent prompt. Repeated Reference
identifiers are aligned independently and inherit the timing of the Shot in
which they occur. Padding, tokenizer special tokens, and connector register
slots receive the sentinel value $(-1,-1)$ and no routing score.

\begin{verbatim}
{
  "references": [{"ref_id": "PERSON_1", ...}],
  "shots": [
    {"shot_id": "SHOT_1", "time_range": [0.0, 2.3],
     "visual_description": "[PERSON_1] turns to the door", ...},
    {"shot_id": "SHOT_2", "time_range": [2.3, 5.0], ...}
  ],
  "events": [
    {"event_id": "DIALOGUE_1", "type": "dialogue",
     "time_range": [1.7, 3.1],
     "content": {"speaker": "PERSON_1", "line": "..."}}
  ],
  "scene_description": "...",
  "global_style": "...",
  "global_audio": "..."
}
\end{verbatim}

For every model, the requested clip duration is set by the final Shot endpoint.
The compiler therefore changes only how script timing enters the conditioning
pathway, not the semantic or temporal information provided.

%

\end{document}